\documentclass[sigchi]{acmart}
\AtBeginDocument{%
  \providecommand\BibTeX{{%
    \normalfont B\kern-0.5em{\scshape i\kern-0.25em b}\kern-0.8em\TeX}}}

\usepackage{outlines}
\usepackage{verbatim}
\usepackage{subcaption}
\usepackage[inline]{enumitem}

\setcopyright{acmlicensed}
\begin{document}
\copyrightyear{2019} 
\acmYear{2019} 
\acmConference[UbiComp/ISWC '19 Adjunct]{Adjunct Proceedings of the 2019 ACM International Joint Conference on Pervasive and Ubiquitous Computing and the 2019 International Symposium on Wearable Computers}{September 9--13, 2019}{London, United Kingdom}
\acmBooktitle{Adjunct Proceedings of the 2019 ACM International Joint Conference on Pervasive and Ubiquitous Computing and the 2019 International Symposium on Wearable Computers (UbiComp/ISWC '19 Adjunct), September 9--13, 2019, London, United Kingdom}
\acmPrice{15.00}
\acmDOI{10.1145/3341162.3344845}
\acmISBN{978-1-4503-6869-8/19/09}
\title[]{The Impact of Private and Work-Related Smartphone Usage on Interruptibility}

\author{Christoph Anderson}
\email{anderson@uni-kassel.de}
\orcid{0000-0002-4082-8457}
\affiliation{%
  \institution{University of Kassel}
  \streetaddress{Wilhelmshoeher Allee 73}
  \city{Kassel}
  \state{Germany}
}

\author{Judith S. Heinisch}
\email{judith.heinisch@uni-kassel.de}
\orcid{}
\affiliation{%
  \institution{University of Kassel}
  \streetaddress{Wilhelmshoeher Allee 73}
  \city{Kassel}
  \state{Germany}
}

\author{Sandra Ohly}
\email{ohly@uni-kassel.de}
\orcid{}
\affiliation{%
  \institution{University of Kassel}
  \streetaddress{Pfannkuchstra\sse 1}
  \city{Kassel}
  \state{Germany}
}

\author{Klaus David}
\email{david@uni-kassel.de}
\orcid{}
\affiliation{%
  \institution{University of Kassel}
  \streetaddress{Wilhelmshoeher Allee 73}
  \city{Kassel}
  \state{Germany}
}

\author{Veljko Pejovic}
\email{veljko.pejovic@fri.uni-lj.si}
\affiliation{%
 \institution{University of Ljubljana}
 \streetaddress{Vecna pot 113}
 \city{Ljubljana}
 \country{Slovenia}}

\renewcommand{\shortauthors}{Anderson et al.}
\begin{abstract}
In the last decade, the effects of interruptions through mobile notifications have been extensively researched in the field of Human-Computer Interaction. Breakpoints in tasks and activities, cognitive load, and personality traits have all been shown to correlate with individuals' interruptibility. However, concepts that explain interruptibility in a broader sense are needed to provide a holistic understanding of its characteristics. In this paper, we build upon the theory of social roles to conceptualize and investigate the correlation between individuals' private and work-related smartphone usage and their interruptibility. Through our preliminary study with four participants over $11$ weeks, we found that application sequences on smartphones correlate with individuals' private and work roles. We observed that participants engaged in these roles tend to follow specific interruptibility strategies -- integrating, combining, or segmenting private and work-related engagements. Understanding these strategies breaks new ground for attention and interruption management systems in ubiquitous computing.
\end{abstract}



\maketitle
\section{Introduction}
Notifications are the leading mechanisms that mobile applications use to engage users with new information proactively. Notifications can inform users about messages, updates, or even recommend content they might be interested in. Despite their practicability, notifications bear the potential to interrupt users in their activities and tasks resulting in stress \cite{Mark:2008}, reduced work performance \cite{Bailey:2006,Leroy:2009}, or uninstalls of irritating applications \cite{Pielot:2017}. Hence, researchers investigated various approaches to intelligently schedule potentially interrupting notifications -- detecting breakpoints in activities \cite{Okoshi:2018} and tasks \cite{Iqbal:2008}, inferring cognitive load from physiological signals \cite{Zuger:2015}, or exploiting personality traits of individuals \cite{Yuan:2017}. However, approaches that consider a broader social context rather than physical or physiological information are needed to provide a holistic view of individuals' interruptibility. 

In this paper, we examine if and how private and work-related smartphone usage influence users' interruptibility. More specifically, we investigate the extent to which private and work-related usage can be identified automatically and if it correlates with users' interruptibility preferences. Through a preliminary study of four participants over $11$ weeks, we show that private and work-related smartphone usage is related to distinctive application patterns. Furthermore, we find indicators that participants follow specific strategies -- integration, combination, or separation -- to manage their interruptibility when being engaged in private or work-related matters.

\section{Social Role Theory \& Interruptibility}
\begin{figure}[tbp]
  \centering
  \includegraphics[width=\linewidth]{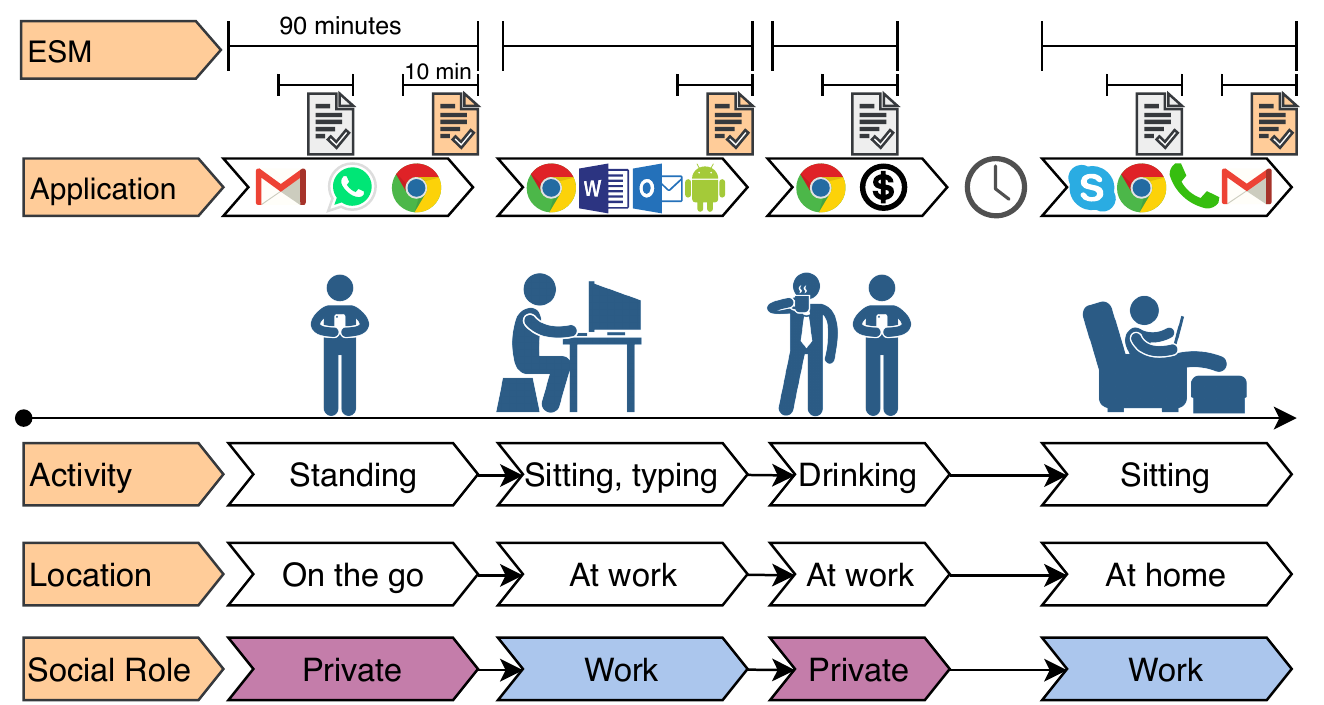}
  \caption{Exemplary illustration of an individual's social roles during the day.}
  \Description{Notification preferences.}
  \label{fig:work_private_roles}
\end{figure}

In this paper, we build upon the theory of \textit{social roles} to conceptualize private and work-related smartphone usage. The theory states that individuals have various roles in life, each associated with distinct norms, behaviors, expectations as well as rights \cite{Biddle:1986}. Figure \ref{fig:work_private_roles} illustrates the basic concepts of the social role theory. While on the go, users might be browsing the internet or chatting with their family and friends, indicating a \textit{private role}. At work, they are engaged in a \textit{work-related role} most of the times. However, they might take a couple of minutes and switch to a \textit{private role}, for example, by checking their private bank accounts using their smartphones. Even though individuals enacting particular roles, sometimes their current role changes, for example, from \textit{work} to \textit{private} and vice versa. Also, individuals not only enact roles, but they also form \textit{spheres} to manage their focus, goals, or styles to fit associated demands. These spheres can be shaped and customized according to individual needs \cite{Clark:2000}. In \cite{Anderson:2016}, we showed different strategies for managing demands in work-private spheres. Individuals who prefer to separate work and private spheres are known as \textit{segmenters}. They try to avoid private and work-related demands while being in work or private-related roles, respectively. In contrast, \textit{integrators} prefer being available for private and work-related demands irrespective of their sphere. At last, \textit{combinators} want to establish a segmentation in one sphere while integrating demands in the other sphere. %
In this paper, we exploit individuals' behaviors and hypothesize that social roles and their associated spheres are reflected by the way individuals use their smartphones. We further suggest that social roles correlate with distinctive interruptibility strategies which might be used in attention and interruption management systems. While the social role theory does not put a limit to the number of roles per se, we restrict our investigation on \textit{work-private} roles for the sake of this study.

\section{Related Work}
A vast amount of literature has investigated the causes and effects of interruptions on individuals \cite{Mehrotra:2015,Pejovic:2014,Fischer:2010,Okoshi:2017}. Mehrotra et al. investigated the concept of association rules for notification preferences~\cite{Mehrotra:2016a}. They found that association rules filtered out about $60 \%$ of unwanted notifications. Furthermore, generated rules were accepted by participants in about $57 \%$ of the cases, indicating the viability of their approach. Gonzalez et al. motivated the use of working spheres -- a concept to describe in which ways individuals organize their work \cite{Gonzalez:2004}. The authors found that individuals spent their time in $10$ working spheres on average. Furthermore, they found that individuals continuously switched between working spheres, partly caused by external interruptions (e.g., colleagues) but also by interrupting themselves through internal interruptions.

The concept of using association rules to learn notification preferences~\cite{Mehrotra:2016a}, as well as considering spheres to describe basic units of work \cite{Gonzalez:2004}, represent the closest works to our approach. Rather than learning rules on an application's level, we focus on determining \textit{social roles} to categorize usage and behavior patterns. The resulting roles and associated spheres are then evaluated to find correlations to interruptibility preferences on a much broader scale. In contrast to \cite{Gonzalez:2004}, where the effects of interruptions for fragmentations of working spheres were examined, we aim to find correlations of \textit{work-private} roles and interruptibility preferences. Nonetheless, Derks et al. showed that work-related smartphone usage during non-office hours might correlate with conflicts of \textit{work-family} roles depending on individual \textit{segmentation} and \textit{integration} strategies \cite{Derks:2016}. However, we investigate how to automatically detect \textit{private-work} roles and interruptibility strategies based on smartphone-usage.     

\section{Study Design}
We developed an Android app to gather events and notifications of individuals (see Table~\ref{tab:data_types}). %
\begin{table}[htbp]
    \centering
	\begin{scriptsize}
	\caption{Data types gathered within the study. $\ast$Manually reported via ESM.}
	\label{tab:data_types}
	\begin{tabular}{p{0.8\linewidth}}
		\toprule                                                  
		\textbf{Notifications}                                              \\
		\midrule                                                  
		Application \{name, package\}										\\
		Notification time \{issued, interacted, posted\}                    \\
		Notification ongoing \{true, false\}                                \\
        \midrule
        \textbf{Contacts}                                                   \\
        \midrule
        Contact \{hash\}                                                    \\
		Relation$\ast$ \{family, friend, work, none\}         	            \\
		\midrule                                                   
		\textbf{Events}                                                    	\\
		\midrule  
		WiFi \{SSID, BSSID\}                                             	\\
		Location \{latitude, longitude\}									\\
		Application \{name, package\}										\\
		Time \{date, issued, interacted, duration\}                         \\
		ESM-Role$\ast$ \{private, work\}                                    \\
		ESM-Interruptibility$\ast$ \{private, work, both, not interruptible\}  \\
		Physical Activity \{Google's Activity Recogn. API\}                 \\
		Power \{connected, disconnected\}                                   \\
		Screen \{on, off\}                                                  \\
		Ringer mode \{silent, vibrate, normal\}                             \\
		\bottomrule                                               
	\end{tabular}
	\end{scriptsize}
\end{table}
Data were collected from four knowledge workers over $11$ weeks. Physical activities and application events were gathered by using Google's Activity Recognition API and the Accessibility Service, respectively. We used experience-sampling methods (ESM), to obtain information about the participants' active social roles \textit{in-situ}. Two ESM questionnaires were issued every $90$ minutes (see Figure~\ref{fig:work_private_roles}, ESM marked in orange): one asking for active social roles within the last $15$ minutes and the other to rate interruptibility preferences. We decided to push additional ESM questionnaires if participants had spent more than $10$ minutes interacting with their smartphone not to miss social role and interruptibility information during active sessions (see Figure~\ref{fig:work_private_roles}, ESM marked in grey). The \textit{ESM-Role} questionnaire let users select their current role -- either \textit{private} or \textit{work}. The \textit{ESM-Interruptibility} questionnaire that followed, allowed the participants to select between being interruptible for only \textit{private} or \textit{work-related} matters, for \textit{both}, or \textit{not being interruptible} at all. Table \ref{tab:data_stats} gives an overview of the gathered data set. In total, $1,574$ ESM-Role questionnaires and $1,047$ ESM-Interruptibility questionnaires had been answered. Furthermore, the application collected $9,943$ notifications and $966,781$ events, thereof $146,783$ application events. At last, participants were regularly asked about the relationship to contacts from whom they received messages. %
\begin{table}[htbp]
	\centering
	\begin{scriptsize}
		\caption{General statistics of the data gathered in our study. $\ast$Contains all events from Table \ref{tab:data_types}. $\dagger$Only answered questionnaires were considered.}
		\label{tab:data_stats}
		\begin{tabular}{p{0.27\linewidth}p{.1\linewidth}p{.1\linewidth}p{.35\linewidth}}
			\toprule  
			\textbf{Description} & \textbf{No.} & \textbf{$\bar No.$} & (\textbf{$\bar P1$/$\bar P2$/$\bar P3$/$\bar P4$})  \\
			\midrule                                                  
			Notifications               & $9,943$     &   $127.4$     & ($33.2/37.7/22.5/36.9$)      \\
			Events$\ast$                & $966,781$   &   $12,394$    & ($3,152/2,357/3,205/3,861$)  \\
			ESM-Role$\dagger$           & $1,574$     &   $20.1$      & ($9.7/4.7/4.2/5.3$)          \\
			ESM-Interruptibility$\dagger$   & $1,047$     &   $13.4$  & ($7.0/3.4/2.6/3.0$)          \\
			\bottomrule                                               
		\end{tabular}
	\end{scriptsize}
\end{table}%
\section{Methodology}
In this paper, we are interested in answering two questions: \begin{enumerate*}[label=\textbf{(\arabic*)}]
  \item Are social roles correlated with individual smartphone usage?
  \item Are social roles, and their associated spheres related to individual interruptibility preferences?
\end{enumerate*} 
To answer the first question, we used the \textit{apriori} algorithm \cite{Agrawal:1993,Agrawal:1996}, to find frequent application sequences that might correlate with the participants' social roles. Let us consider the following example to illustrate the principles of this algorithm. Let $T = \{T_1, T_2,...,T_n\}$ be a multiset of transactions, where each transaction $T_i \in T$ is a set of applications that has been used and labeled by an ESM questionnaire within the last $15$ minutes. The algorithm then finds frequent application sets $S$ in $T$ with a given \textit{support}. The support of an application set $S_i \in S$ is defined as the minimum number of transactions $T_i$ that contain $S_i$. The following example shows the computation of frequent application sets with $support=2$.
\begin{equation*}
\label{eq:transaction}
    \begin{aligned}
    T &= \{T_1, T_2\} \\
    T_1 &= \{Whatsapp, Paypal, Gmail, Teams\} \\
    T_2 &= \{Phone, Paypal, Whatsapp, Reddit\} \\
    S &= \{\{Whatsapp\},\{Paypal\}, \{Whatsapp, Paypal\}\}
    \end{aligned}
\end{equation*}
It is worth noting that the algorithm does not consider the order nor multiple occurrences of applications within sequences. The second question, whether social roles (independent variable) are related to interruptibility preferences (depended variable), is answered by applying multiple significance tests. Since our data contain categorical values, we chose the \textit{Chi-Square test of independence} \cite{McHugh:2013} to examine if a statistically significant relationship between social roles and interruptibility preferences exists. For a more detailed analysis, we binary encoded ESM questionnaires in either being interruptible for \textit{private} or for \textit{work-related} matters using the schema described in Table \ref{tab:binary_encoding}. The resulting encoding is then used to investigate potential individual preferences for each social role.
\begin{table}[h]
    \centering
    \begin{scriptsize}
    \caption{Binary encoding of interruptibility preferences}
    \begin{tabular}{c|c|c}
        \toprule
        \textbf{Preference} & \textbf{Private} & \textbf{Work}  \\
        \midrule
        none    &   0   &   0   \\
        work    &   0   &   1   \\
        private &   1   &   0   \\
        both    &   1   &   1   \\
        \bottomrule
    \end{tabular}
    \label{tab:binary_encoding}
    \end{scriptsize}
\end{table}%

\section{Results}
In this section, we present the results on the identification of social roles and interruptibility strategies based on our preliminary study. We first analyze frequent application sequences to determine participant's social roles. We then present and discuss the results of interruptibility strategies in relation to social roles. 

\subsection{Determination of Social Roles By Application Usage}
\begin{figure}[t]
\centering
    \begin{subfigure}[a]{.7\linewidth}
        \includegraphics[width=.49\linewidth]{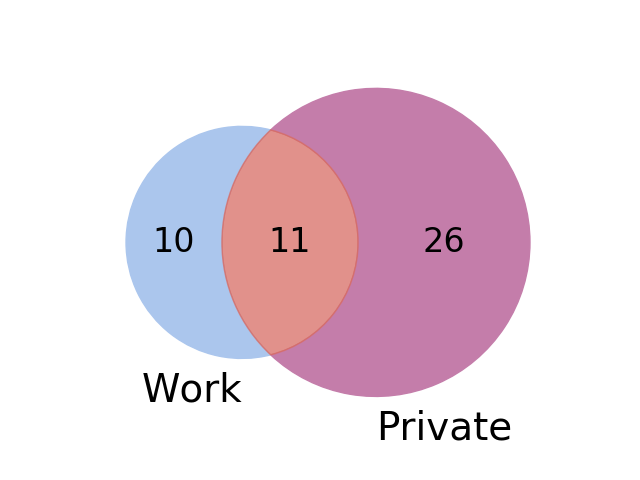}
        \includegraphics[width=.49\linewidth]{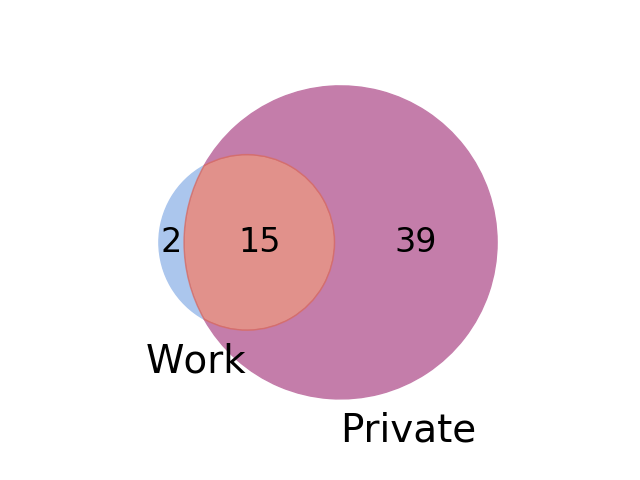}
        \caption{Sequences P1 (left), P2 (right)}
        \label{fig:sequences_a}
    \end{subfigure}
    \begin{subfigure}[a]{.7\linewidth}
        \includegraphics[width=.49\linewidth]{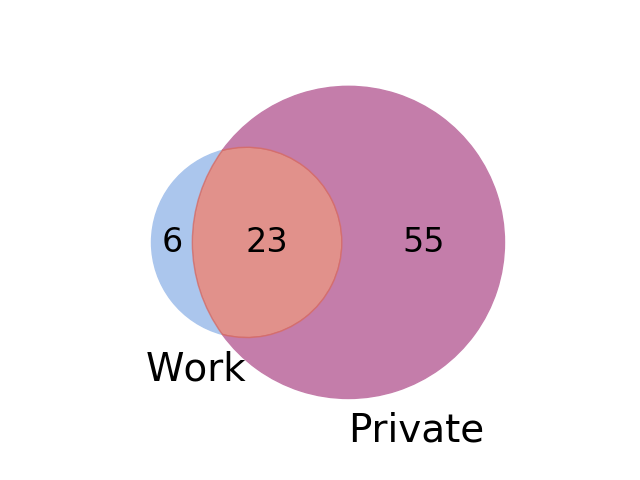}
        \includegraphics[width=.49\linewidth]{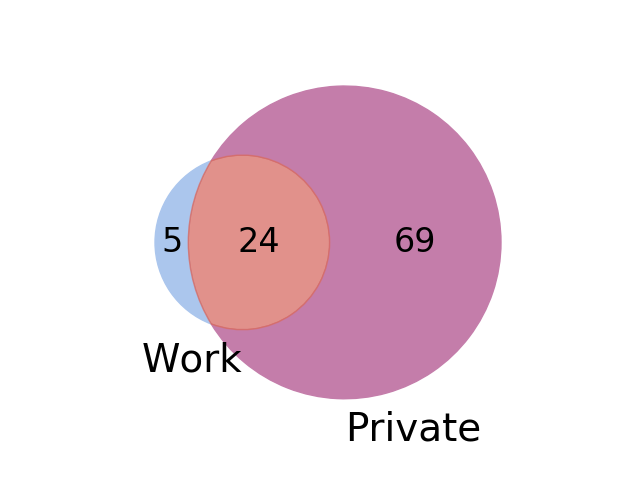}
        \caption{Sequences P3 (left), P4 (right)}
         \label{fig:sequences_b}
    \end{subfigure}
  \caption{Frequent application sequences for private and work roles}
  \label{fig:sequences}
\end{figure}

We noticed that our participants more frequently reported their social role as being \textit{private} rather than being \textit{work}. We also found that the number of distinct applications used while being \textit{private} was greater than the number of distinct applications used while being engaged in \textit{work}. This was true for all participants. Figure \ref{fig:sequences} illustrates the sequence analysis of applications used in private and work roles using a support-value of $5 \%$. The analysis shows that predominantly unique private sequences and less explicit work-related sequences were found in the individual participants' transactions. A more detailed analysis of the sequences suggests that participants, to some extent, use individual applications in private and work-related roles. For example, the following exclusively work-related sequences were found for participant P1:  
\begin{quote}
    \{\{Gboard, Slack\}, \{System-UI, Teams\}, \{Slack\}, \{System-UI, Settings\}, \{Gboard, System-UI, Teams\}, \{Gboard, Teams\}, \{System-UI, Slack\}, \{Settings\}, \{Teams\}, \{Gboard, System-UI, Slack\}\}
\end{quote}%
Participant P1 used \textit{Slack} and \textit{Teams} for work-related purposes, whereas \textit{Gboard} and \textit{System-UI} were also found in private sequences. This seems obvious as the former is a keyboard application and the latter is recorded when interacting with the notification bar. Sequences containing either \textit{Teams} or \textit{Slack} were also observed for participant P3 and P4 but not for P2. In return, none of P2's work-related sequences were found in the respective sequences of the other participants. Similarly, the same observation applies to private sequences, for which individual applications (e.g., Chess, Geocaching) were found as well.
We also observed a significant overlap between private and work-related sequences, which could neither be explicitly assigned to one nor the other role. We noticed that browsers and messengers frequently appeared in these intersection sequences. Further analysis of the contacts the participants communicated with showed that messengers, such as \textit{Whatsapp} or \textit{Threema} were not only used to chat with friends and family members, but also with colleagues at the same time. Although participants received messages from private and work-related contacts, messengers such as \textit{Teams} were only found in work associated sequences. This finding might indicate a distinction of messengers according to work-private related purposes.

\subsection{Analysis \& Significance of Interruptibility Strategies}
We computed a Chi-Square test of independence on the reported social roles and interruptibility preferences (see Table \ref{tab:significance_results}). The test revealed that there is a significant statistical dependency between social roles and interruptibility preferences $\chi^{2}(3, N = 4) = 382.07, p < 0.01$. Further analysis showed a dependency between social roles and interruptibility preferences for all participants, setting $\alpha$ as $0.01$.
\begin{table}[htbp]
	\centering
	\begin{scriptsize}
		\caption{$\chi^{2}$-Square test of independence results for social roles and interruptibility preferences.}
		\label{tab:significance_results}
		\begin{tabular}{p{.2\linewidth}p{.2\linewidth}p{.2\linewidth}p{.2\linewidth}}
			\toprule                                                  
			\textbf{Participant} & \textbf{$\chi^{2}$} & \textbf{$p-value$} & DOF \\
			\midrule
			$P1$       & $243.18$  & $1.95e-52$  & $3$  \\
			$P2$       & $127.79$  & $1.61e-27$  & $3$  \\
			$P3$       & $16.46$   & $0.00091$   & $3$  \\
			$P4$       & $13.94$   & $0.00018$   & $1$  \\
			\midrule  
			$All$      & $382.07$  & $1.69e-82$  & $3$  \\
			\bottomrule                                               
		\end{tabular}
	\end{scriptsize}
\end{table}%
\begin{figure*}[t]
\centering
    \begin{subfigure}[c]{.49\linewidth}
    \centering
        \includegraphics[scale=.35]{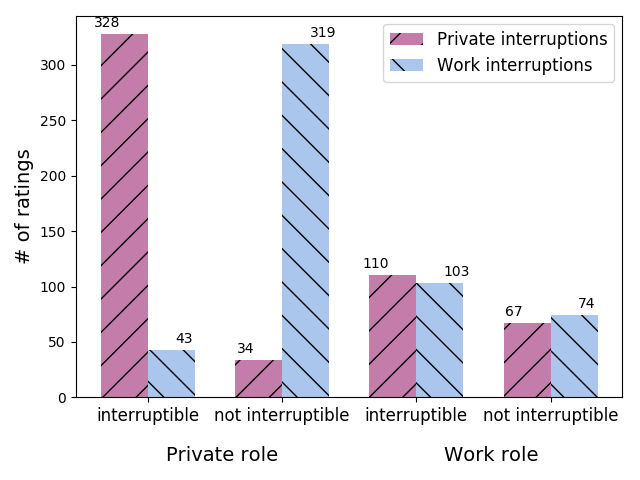}
        \subcaption{Participant P1}
        \label{fig:preferences_p1}
    \end{subfigure}
    \begin{subfigure}[c]{.49\linewidth}
    \centering
        \includegraphics[scale=.35]{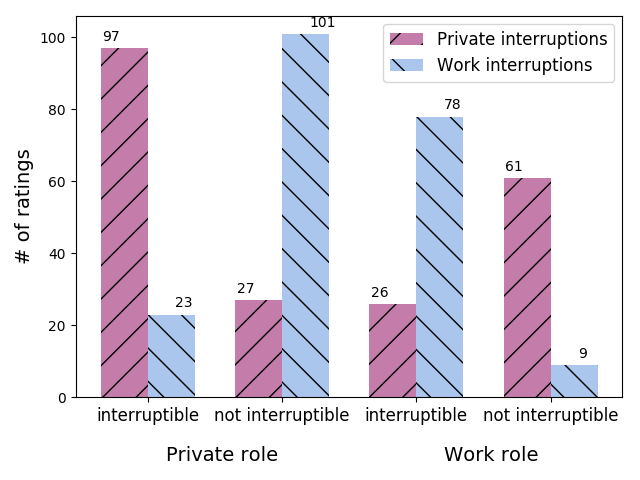}
        \subcaption{Participant P2}
        \label{fig:preferences_p2}
    \end{subfigure}
    \begin{subfigure}[c]{.49\linewidth}
    \centering
        \includegraphics[scale=.35]{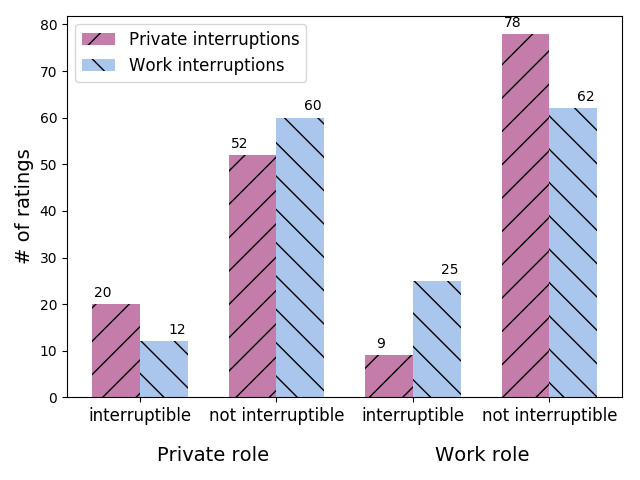}
        \subcaption{Participant P3}
        \label{fig:preferences_p3}
    \end{subfigure}
    \begin{subfigure}[c]{.49\linewidth}
    \centering
        \includegraphics[scale=.35]{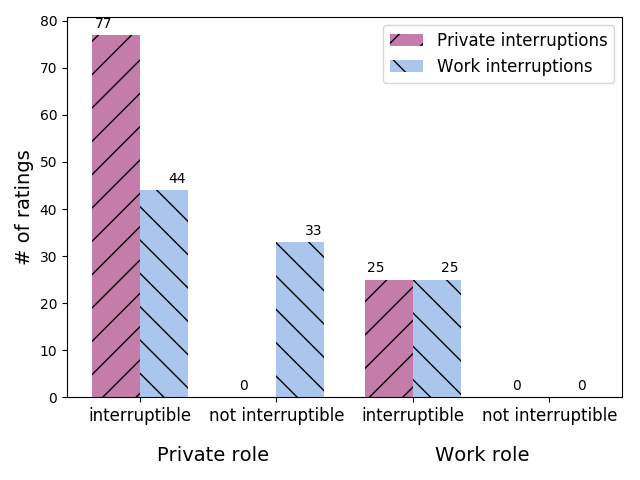}
        \subcaption{Participant P4}
        \label{fig:preferences_p4}
    \end{subfigure}
    \caption{Interruptibility ratings in relation to private and work roles.}
    \label{fig:preferences}
\end{figure*}
We further analyzed binary encoded ESM interruptibility questionnaires -- investigating if there is a specific dependency between social roles and private and work-related interruptibility preferences. The results of the participants' binary encoded interruptibility preferences are shown in Figure \ref{fig:preferences} and in Table \ref{tab:binary_significance_results}. Considering the private role, the interruptibility ratings of participant P1 and P2 suggest a segmentation preference (see Figures \ref{fig:preferences_p1} - \ref{fig:preferences_p2}). In particular, participants P1 and P2 were more likely interruptible for private but less likely interruptible for work-related demands while enacting a private role. However, the interruptibility ratings for both participants differentiated when enacting a work role. Whereas participant P1 showed an integration preference demonstrated by an almost equal distribution of ratings, participant P2 seemed to prefer a segmentation preference when enacting a work role. Considering interruptibility ratings for both roles, these findings suggest that participant P1 might follow a \textit{combination} strategy -- being interruptible for private and work-related matters in the work role but \textit{not} interruptible for work-related demands while being private. The preferences of participant P1 are significantly dependent to the social role -- $\chi^{2}(1, N = 1) = 61.38, p < 0.01$ and $\chi^{2}(1, N = 1) = 126.77, p < 0.01$, respectively. The overall ratings of participant P2 indicate a \textit{segmentation} strategy -- being \textit{only} interruptible for private matters in the private role and for work-related demands in the work-role, respectively. The preferences of participant P2 are significantly dependent on the social role -- $\chi^{2}(1,N = 1) = 47.17, p < 0.01$ and $\chi^{2}(1, N = 1) = 100.76, p < 0.01$, respectively.
\begin{table}[htbp]
	\centering
	\begin{scriptsize}
		\caption{$\chi^{2}$-Square test of independence results for social roles and binary encoded interruptibility preferences.}
		\label{tab:binary_significance_results}
		\begin{tabular}{p{.2\linewidth}p{.2\linewidth}p{.2\linewidth}p{.2\linewidth}}
			\toprule                                                  
			\textbf{Participant} & \textbf{$\chi^{2}$} & \textbf{$p-value$} & DOF \\ 
			\midrule
			$P1_{private}$       & $61.38$   & $4.70e-15$  & $1$    \\
			$P1_{work}$          & $126.77$  & $2.08e-29$  & $1$    \\
			\midrule
			$P2_{private}$       & $47.17$   & $6.49e-12$  & $1$    \\
			$P2_{work}$          & $100.76$  & $1.03e-23$  & $1$    \\
			\midrule
			$P3_{private}$       & $6.90$   & $0.0086$  &   $1$     \\
			$P3_{work}$          & $2.57$   & $0.10$    &   $1$     \\
			\midrule
			$P4_{private}$       & $-$      & $-$       &   $-$     \\
			$P4_{work}$          & $-$      & $-$       &   $-$     \\
			\midrule
			$All_{private}$      & $147.92$ & $4.92e-34$ & $1$      \\
			$All_{work}$         & $183.42$ & $8.66e-42$ & $1$      \\
			\bottomrule                                               
		\end{tabular}
	\end{scriptsize}
\end{table}
Indicators for a potential \textit{integration} strategy were observed for participant P3. Note that \textit{integrators} prefer being available for private and work-related demands irrespective of their role. Hence, there might be no dependency between social roles and their interruptibility preferences. The Chi-Square test of independence seems to confirm this suggestion. When enacting a work role, \textit{no} significant correlation between social roles and work-related interruptibility preferences was found $\chi^{2}(1, N = 1) = 2.57, p > 0.01$. However, a weak dependency was found for private-related interruptibility preferences and social roles $\chi^{2}(1, N = 1) = 6.90, p < 0.01$. A detailed analysis revealed that participant P3 frequently answered as not being interruptible at all, suggesting that the results might be ambiguous. Participant P4 seemed to follow a \textit{combination} strategy. However, the results might also be less univocal as the participant only reported to be interruptible for \textit{both} or only for \textit{private matters}. Therefore, we could not perform a significance test on the interruptibility preferences of participant P4 due to insufficient data.

\section{Limitations \& Future Work}
In this paper, we investigate the correlation between private and work-related smartphone usage and interruptibility. We find indicators that help us to determine an user's social roles by investigating frequent application sequences. Furthermore, our findings suggest that users tend to follow specific interruptibility strategies while enacting private and work roles. However, the research findings in this study are limited due to the following reasons. First, only smartphone usage data was considered to examine social roles and interruptibility. As the participants primarily worked with laptops or computers, a substantial amount of work-related usage data was not collected and, therefore, not analyzed. Secondly, the relevance of our analysis is limited, as only four participants were examined in our preliminary study. We plan to mitigate these limitations by implementing a desktop application that gathers relevant data as well as by increasing the number of participants. Furthermore, we aim to better distinguish application sequences at the intersection of private and work roles as well as to investigate more diverse social roles. We are certain to confirm our preliminary results in future and larger studies.

\section{Conclusion}
In this paper, we show that social roles associated with distinct norms, behaviors, expectations as well as rights are correlated with individuals' interruptibility preferences. Our study reveals that when individuals enact a social role, they tend to follow specific strategies to manage their interruptibility. For our analysis, we first collected data of smartphone-usage and notification preferences in an $11$-week \textit{in-the-wild} study with four participants. Through a sequential analysis of smartphone applications, we found distinctive and unique sequences associated with individuals' private and work roles. A statistical significance between social roles and rated interruptibility preferences was found by applying multiple Chi-Square tests of independence. Furthermore, a binary encoding of interruptibility preferences revealed three types of interruptibility strategies - segmentation, integration, and a combination of private and work-related engagements. Understanding these strategies breaks new ground for attention and interruption management systems in ubiquitous computing. In future, users might be supported in their interruptibility by postponing notifications according to individual strategies. 
\begin{acks}
The authors would like to thank Slaven Bogdanovi\'{c}, Iris \u{Z}e\u{z}elj, and Daniel Wilhelm for their fruitful discussions and comments that greatly improved this paper.
\end{acks}
\bibliographystyle{ACM-Reference-Format}
\bibliography{bibliography}

\end{document}